\documentclass{elsarticle}
\usepackage[cp1251]{inputenc}
\usepackage{amssymb,amsmath}
\usepackage{textcomp}
\usepackage{graphicx}
\usepackage[all]{xy}
\usepackage[usenames]{color}
\usepackage{colortbl}

\usepackage{hhline}
\usepackage{longtable}
\usepackage{fancyvrb}

\journal{   }

\begin{document}

\begin{frontmatter}

\title{Secure pseudo-random linear binary sequences generators based on arithmetic polynoms}
%

\author[ITP]{Oleg Finko}
\ead{ofinko@yandex.ru}

\author{Sergey Dichenko}
\ead{dichenko.sa@yandex.ru}

\address[ITP]{Computer Systems and Information Security of KubSTU,
Krasnodar, Russia}


\begin{abstract}
We present a new approach to constructing of pseudo-random binary sequences
(PRS) generators for the purpose of cryptographic data protection, secured
from the perpetrator's attacks, caused by generation of masses of hardware
errors and faults. The new method is based on use of linear polynomial
arithmetic for the realization of systems of boolean characteristic functions
of PRS' generators. ``Arithmetizatio'' of systems of
logic formulas has allowed to apply mathematical apparatus of residue systems
for multisequencing of the process of PRS generation and organizing control
of computing errors, caused by hardware faults. This has guaranteed high
security of PRS generator's functioning and, consequently, security of
tools for cryptographic data protection based on those PRSs.
\end{abstract}
%

\end{frontmatter}

\section{Introduction}
PRS' generators play an important role in building
of communication with cryptographic data protection [1, 2]. From the list of
known attacks on information security is important type of attacks, based on
the generation of hardware errors functioning of the nodes forming the binary
PRS [3]. To ensure the required level of interference and fault tolerance of
digital devices developed many methods, the most common of which are backup
methods and methods of error-correcting coding [4]. However, allocation methods
do not provide the required levels of fault tolerance for restrictions on
hardware costs, and methods of error-correcting coding is not adapted to the
specifics of construction and operation means of data protection (MDP),
in particular, the generators of the PRS.

\section{Analysis of attacks based on hardware faults generation}

Currently, the following types of attacks on sites of formation of
binary PRS are considered (attack on) [5]:
\begin{itemize}
    \item analysis of results of power consumption measurements;
    \item analysis of results of operations performance duration;
    \item analysis of accidental hardware faults;
    \item analysis of intentionally generated hardware faults, etc.
\end{itemize}

The last two types of faults are not investigated enough currently
and thus are threatening to the information security of the functioning
of modern and perspective MDP. The origin of those attacks lies in the
use of thermal, high frequency, ionizing and other types of external
influences onto MDP for the purpose of creation of masses of faults
in hardware functioning by initialising of computing errors.

Hardware attacks can be divided into two classes:
\begin{enumerate}
     \item \textbf{Direct hardware attacks.} The consequences of those
     attacks are failures of data protection tools. There is a method
     of analysis of the consequences of those failures. These types of
     attacks mean that in distortion in the certain places algorithm of
     transformation, which results in computing errors. Those errors can
     lead, for example, to repeated generation of the elements of PRS or
     in generation of faulty elements of PRS, which is unacceptable
    \item \textbf{Attacks on post failure recovery means.} Some systems
    do not recovery means. If the system protection is destroyed, it is
    impossible to restore the operational mode. That is why such systems
    need to have means of protection against attacks of the malefactor
    and to support the possibility of updating the security system without
    stopping the programme running.
\end{enumerate}

Attacks, based on errors generation by means of external influence are
highly efficient for the majority of currently known and used algorithms
of PRS generation. It is known that probability of error generation is
proportional to the time corresponding registers has been affected by
the radiation, if the registers are in favourable condition for error
occurrence, and to the quantity of bits, in which the error occurrence
is expected. The most widely used and proven means of creating
PRS are algorithms and structures~--- Linear feedback shift register
(LFSR)~--- of PRS generation, based on the use of feedback functions
of logic [1, 2].

The structure of LFSR is determined by the forming polynomial:
\begin{align*}
    D(\chi)=\chi^\tau+\chi^{t_{l}}+\ldots+\chi^{t_{2}}+\chi^{t_{1}}+1,
\end{align*}
where $\tau,\ t_{i}\in N$ and characteristic equation based on it:
\begin{align}
    \begin{split}
        x_{p+\tau}&=x_{p}\oplus x_{p+t_{1}}\oplus x_{p+t_{2}}\oplus\ldots\oplus x_{p+t_{l}} \\
        &=c_{0}x_{p}\oplus c_{1}x_{p+1}\oplus \ldots\oplus c_{\tau-2}x_{p+\tau-2}\oplus c_{\tau-1}x_{p+\tau-1},
    \end{split}
\end{align}
where $x_{p},\ c_{i}\in \{0,\ 1\}$;  $p\in N$;  $i=0,\ 1,\
\ldots,\ \tau-1$; $c_{i\in \{0,\ t_{1},\ t_{2},\ \ldots,\
t_{l}\}}=1$.

In linear algebra the next element of PRS $x_{p+\tau}$
is calculated as the following multiplication:
\begin{align*}
    \begin{Vmatrix}
        x_{p+1} \\
        x_{p+2} \\
       \cdots\cdots \\
        x_{p+\tau-1} \\
        x_{p+\tau}
    \end{Vmatrix}
    ^{\top}=
    \begin{Vmatrix}
        0 & 1 & \ldots\ldots & 0 & 0 \\
        \hdotsfor[2]{5} \\
        0 & 0 & \ldots\ldots & 1 & 0 \\
        0 & 0 & \ldots\ldots & 0 & 1 \\
        c_{0} & c_{1} & \ldots\ldots & c_{\tau-2} & c_{\tau-1}
    \end{Vmatrix}^{\top}
    \cdot
    \begin{Vmatrix}
        x_{p} \\
        x_{p+1} \\
         \cdots\cdots \\
        x_{p+\tau-2} \\
        x_{p+\tau-1}
    \end{Vmatrix}^{\top}.
\end{align*}

When the described attack is performed the conditions arise
for PRS modification or its repeated generation.
The effect of repeated generation of a site of PRS is explained by
means of Fig.~1 (the forming polynomial: $D(\chi)=\chi^4+\chi+1$;
the characteristic equation: $x_{p+4}=x_{p+1}\oplus x_{p}$; the initial conditions: $x_{p}=1$, $x_{p+1}=0$, $x_{p+2}=1$, $x_{p+3}=0$ ).

\begin{figure}[h!]
    \begin{center}
    \resizebox{0.72\linewidth}{!}{
\colorbox[gray]{.92}{
$$
\xymatrix{
 \text{\textnumero} & x_{p+3}               & x_{p+2}                                         & x_{p+1}                                      &                                                                                                 & x_{p}                                                                           & \text{\textbf{\textsf{Error PRS}}}           & \text{\textbf{\textsf{Correct PRS}}}\\
\textbf{\textsf{0}} & 0                              & 1                                                   & 0  \ar@{->}[r]                             & \bigoplus                                                                                       &\ar@{->}[l]   1  \ar@{->}[r] ^{\text{\textsf{Output}}}       &            1\ar@{<-->}[r] &1\\
\textbf{\textsf{1}} & 1                                     & 0                                                   & 1 \ar@{->}[r]                             & \ar@{-->}[llld]|{\text{\textsf{Error (inversion)}}} \bigoplus &   0  \ar@{->}[l]\ar@{->}[r] ^{\text{\textsf{Output}}}                 &                          0\ar@{<-->}[r]&0\\
\textsf{\textbf{2}} & \boxed{\neg 1} \ar@{-->}[rd] & 1                                                    & 0 \ar@{->}[r]                             &  \bigoplus                                                                                     &  1 \ar@{->}[l]  \ar@{->}[r] ^{\text{\textsf{Output}}}       &                   1\ar@{<-->}[r]&1\\
\textsf{\textbf{3}} & 1                                                   & \boxed{\neg{1}}\ar@{-->}[rd] & 1 \ar@{->}[r]                               & \bigoplus                                                                                      &  0 \ar@{->}[l]  \ar@{->}[r] ^{\text{\textsf{Output}}}      &                  0\ar@{<-->}[r]&0\\
\textbf{\textsf{4 }}&                             1                       &                                                 1 & \boxed{\neg{1}}\ar@{-->}[rrd]\ar@{->}[r] & \ar@{-->}[llld] \bigoplus &  1 \ar@{->}[l] \ar@{->}[r] ^{\text{\textsf{Output}}}  &                             1\ar@{<-->}[r]&1   \\
\textbf{\textsf{5}} &  \boxed{\neg{0}}\ar@{-->}[rd]  &                                                 1 & 1\ar@{->}[r]                               &  \ar@{-->}[llld]                    \bigoplus  & \ar@{->}[l]\boxed{\neg{1}} \ar@{->}[r] ^{\text{\textsf{Output}}} &          0\ar@{<-->}[uur]    &1\\
\textbf{\textsf{6}} &  \boxed{\neg{0}}\ar@{-->}[rd]  &\boxed{\neg{0}}\ar@{-->}[rd] & \ar@{->}[r]                                 1& \bigoplus                                                                                      & 1\ar@{->}[l] \ar@{->}[r] ^{\text{\textsf{Output}}}            &  \ar@{<-->}[uur]  1&1\\
\textbf{\textsf{7 }}&                                                    0&\boxed{\neg{0}}\ar@{-->}[rd] & \ar@{->}[r]\boxed{\neg{0}}\ar@{-->}[rrd]    & \ar@{-->}[llld] \bigoplus                                               & 1\ar@{->}[l]\ar@{->}[r] ^{\text{\textsf{Output}}}           &  \ar@{<-->}[uur]   1&1\\
\textbf{\textsf{8}} &  \boxed{\neg{1}} \ar@{-->}[rd]    &                      0& \ar@{->}[r]\boxed{\neg{0}}\ar@{-->}[rrd] &\ar@{-->}[llld] \bigoplus                                                 & \boxed{\neg{0}}\ar@{->}[l]\ar@{->}[r] ^{\text{\textsf{Output}}}           &   \ar@{<-->}[uur]  1       &0\\
\textbf{\textsf{9}} &   \ar@{-->}[rd]         0&\boxed{\neg{1}} \ar@{-->}[rd]  & \ar@{->}[r]                                   0&\ar@{-->}[llld] \bigoplus                                                                &\boxed{\neg{0}}\ar@{->}[l]\ar@{->}[r] ^{\text{\textsf{Output}}}           &  \ar@{<-->}[uur]  1    &0\\
\textbf{\textsf{10 }}& \boxed{\neg{0}}\ar@{-->}[rd]     &    \ar@{-->}[rd]                        0& \ar@{->}[r]\boxed{\neg{1}}\ar@{-->}[rrd]  & \ar@{-->}[llld]\bigoplus                                                   & 0\ar@{->}[l]\ar@{->}[r] ^{\text{\textsf{Output}}}           & \ar@{<-->}[uur]   0&0\\
\textsf{\textbf{11}} & \boxed{\neg{1}} \ar@{-->}[rd]   &\boxed{\neg{0}}\ar@{-->}[rd]      & \ar@{->}[r]  \ar@{-->}[rrd]                0   & \ar@{-->}[llld] \bigoplus                                      & \boxed{\neg{1}} \ar@{->}[l]\ar@{->}[r] ^{\text{\textsf{Output}}}  & \ar@{<-->}[uur] 0 &1\\
\textsf{\textbf{12}} &  \ar@{-->}[rd]  \boxed{\neg{1}}& \boxed{\neg{1}} \ar@{-->}[rd] & \ar@{->}[r]\boxed{\neg{0}}\ar@{-->}[rrd] & \bigoplus                                                                            & 0\ar@{->}[l]\ar@{->}[r] ^{\text{\textsf{Output}}}             & \ar@{<-->}[uur] 0&0\\
\textsf{\textbf{13}} &                          0& \ar@{-->}[rd]  \boxed{\neg{1}}& \ar@{->}[r]\boxed{\neg{1}} \ar@{-->}[rrd] & \ar@{-->}[llld] \bigoplus                                                   & \boxed{\neg{0}}\ar@{->}[l]\ar@{->}[r] ^{\text{\textsf{Output}}}        & \ar@{<-->}[uur]  1&0\\
\textsf{\textbf{14}} &  \ar@{-->}[rd]           1&                                 0& \ar@{->}[r] \ar@{-->}[rrd] \boxed{\neg{1}}& \ar@{-->}[llld]  \bigoplus            & \boxed{\neg{1}} \ar@{->}[l]\ar@{->}[r] ^{\text{\textsf{Output}}}      & \ar@{<-->}[uur]  0   &1 \ar@{-->}[uuuuuuuuuuuuuul]  \\
\textsf{\textbf{15}} &   \boxed{\neg{0}} \ar@{-->}[rd] & \ar@{-->}[rd]                          1& \ar@{->}[r]  0& \ar@{-->}[llld] \bigoplus  & \boxed{\neg{1}}\ar@{->}[l]\ar@{->}[r] ^{\text{\textsf{Output}}}&  \ar@{<-->}[uur]   0&  1\\
\textsf{\textbf{16 }}&  \boxed{\neg{1}}                       & \boxed{\neg{0}}                         & \ar@{->}[r] 1                                    & \bigoplus                                     & 0\ar@{->}[l]\ar@{->}[r] ^{\text{\textsf{Output}}}  &  \ar@{<-->}[uur] 0&  0
}
$$
}
}
  \end{center}
  \caption{Example of operation of the LFSR when an error occurs ($\neg{x}$~--- logical inversion $x$)}
        \end{figure}

Thus, those attacks, which are based on creating the conditions under
which mass hardware errors occur, are threatening for MDP. One of the
ways of solving this problem is development of methods for increasing
the reliability of the functioning of sites of data protection tools,
mostly subjected to attacks oft he described type, in particular the
sites of forming of the encryption algorithm (cipher), based on PRS
generation.

\section{Analysis of methods for reliable binary PRS generation}

Currently the required level of functional reliability of the sites
of binary PRS generation is reached both by using of excessive devices
(reservation) and timely excess by various repetitions of the calculations.
In digital schemotechnics there are solutions known, based on use of
methods of error-correction coding [4]. In order to use those methods
for PRS generators it is necessary preliminary to solve the issue
multisequencing the process of PRS calculations. The solution is based
on the use of classic parallel algorithms of recursion [10].

For example, for the characteristic equation:
\begin{align}
    x_{p+\tau}=x_{p+t}\oplus x_{p},
\end{align}
corresponding to treen $D(\chi)=\chi^\tau+\chi^{t}+1$,
 it is possible to build a system of characteristic equations:
\begin{align*}
    \begin{cases}
        x_{q,\ \tau-1}=x_{q-1,\ \tau-1}\oplus x_{q-1,\ \tau+t-1},\\
        x_{q,\ \tau-2}=x_{q-1,\ \tau-2}\oplus x_{q-1,\ \tau+t-2},\\
        \cdots\cdots\cdots\cdots\cdots\cdots\cdots\cdots\cdots\cdots\cdots\\
        x_{q,\ 1}=x_{q-1,\ 1}\oplus x_{q-1,\ t+1},\\
        x_{q,\ 0}=x_{q-1,\ 0}\oplus x_{q-1,\ t}.
    \end{cases}
\end{align*}

Similarly, for the general equation (1):
\begin{align}
    \begin{cases}
        x_{q,\ \tau-1}=c^{(\tau-1)}_{0}x_{q-1,\ 0}\oplus c^{(\tau-1)}_{1}x_{q-1,\ 1}\oplus \ldots \oplus c^{(\tau-1)}_{\tau-2}x_{q-1,\ \tau-2}\oplus c^{(\tau-1)}_{\tau-1}x_{q-1,\ \tau-1},\\
        x_{q,\ \tau-2}=c^{(\tau-2)}_{0}x_{q-1,\ 0}\oplus c^{(\tau-2)}_{1}x_{q-1,\ 1}\oplus \ldots \oplus c^{(\tau-2)}_{\tau-2}x_{q-1,\ \tau-2}\oplus c^{(\tau-2)}_{\tau-1}x_{q-1,\ \tau-1},\\
        \cdots\cdots\cdots\cdots\cdots\cdots\cdots\cdots\cdots\cdots\cdots\cdots\cdots\cdots\cdots\cdots\cdots\cdots\cdots\cdots\cdots\cdots
        \cdots\cdots\cdots\cdots\\
        x_{q,\ 1}=c^{(1)}_{0}x_{q-1,\ 0}\oplus c^{(1)}_{1}x_{q-1,\ 1}\oplus \ldots \oplus c^{(1)}_{\tau-2}x_{q-1,\ \tau-2}\oplus c^{(1)}_{\tau-1}x_{q-1,\ \tau-1},\\
        x_{q,\ 0}=c^{(0)}_{0}x_{q-1,\ 0}\oplus c^{(0)}_{1}x_{q-1,\ 1}\oplus \ldots \oplus c^{(0)}_{\tau-2}x_{q-1,\ \tau-2}\oplus c^{(0)}_{\tau-1}x_{q-1,\ \tau-1},
    \end{cases}
\end{align}
where $c^{(j)}_{i}\in \{0,\ 1\}$ $(i,\  j=0,\ 1,\ \ldots,\ \tau-1)$. The principle of parallel lasing elements PRS based on (3) is illustrated by a graph (see Fig. 2)

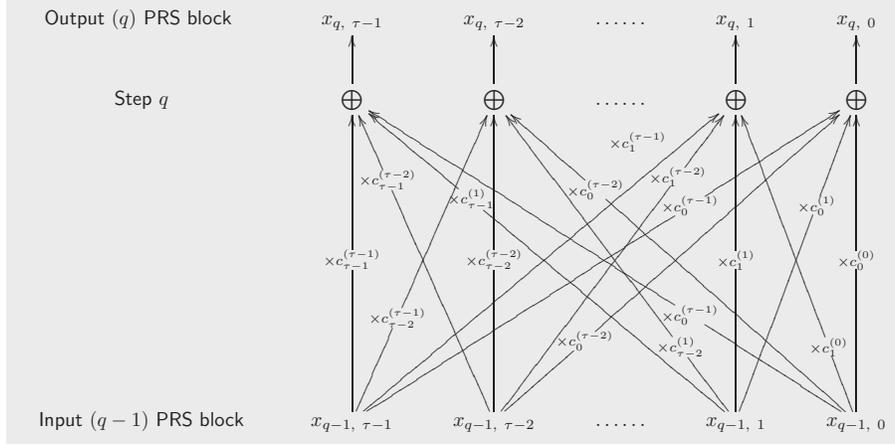
\begin{figure}[h]
    \begin{center}
    \resizebox{1.0\linewidth}{!}{
\colorbox[gray]{.92}{
$$
\xymatrix{
\text{\textsf{Output $(q)$ PRS block} }    &    x_{q,\ \tau -1}                                           & x_{q,\ \tau -2}                                            & \ldots\ldots                            & x_{q,\ 1}                                                                & x_{q,\ 0}    \\
\text{\textsf{Step}\ $q$}                  & \bigoplus   \ar@{->}[u]                          & \bigoplus  \ar@{->}[u]                       & \ldots\ldots                        & \bigoplus   \ar@{->}[u]                                              & \bigoplus  \ar@{->}[u]                               \\
&&&&&\\
&&&&&\\
&&&&&\\
&&&&&\\
\text{ \textsf{Input $(q-1)$  PRS block} } & x_{q-1,\ \tau -1}
\ar@{->}[uuuuu]|{\times c^{(\tau -1)}_{\tau -1}}
\ar@{->}[uuuuur]|(.32){\times c^{(\tau -1)}_{\tau
-2}}\ar@{->}[uuuuurrr]^(.82){\times c^{(\tau
-1)}_{1}}\ar@{->}[uuuuurrrr]|(.67){\times c^{(\tau -1)}_{0}}    &
x_{q-1,\ \tau -2}\ar@{->}[uuuuu]|{\times c^{(\tau -2)}_{\tau
-2}} \ar@{->}[uuuuul]|(.75){\times c^{(\tau -2)}_{\tau -1}}
\ar@{->}[uuuuurr]|(.76){\times c^{(\tau
-2)}_{1}}\ar@{->}[uuuuurrr]|(.25){\times c^{(\tau -2)}_{0}}  &
\ldots\ldots     & x_{q-1,\ 1}\ar@{->}[uuuuu]|{\times c^{(1)}_{1}}
\ar@{->}[uuuuull]|(.23){\times c^{(1)}_{\tau -2}}
\ar@{->}[uuuuulll]|(.69){\times c^{(1)}_{\tau -1}}
\ar@{->}[uuuuur]|(.67){\times c^{(1)}_{0}}       &  x_{q-1,\ 0}
\ar@{->}[uuuuu]|{\times c^{(0)}_{0}}  \ar@{->}[uuuuul]|(.23){\times
c^{(0)}_{1}}   \ar@{->}[uuuuulll]|(.72){\times c^{(\tau -2)}_{0}}
\ar@{->}[uuuuullll]|(.33){\times c^{(\tau -1)}_{0}}
 }
$$
}
}
  \end{center}
 \caption{Graph generating elements parallel PRS based on (3)}
        \end{figure}

System (3) forms an information matrix:
\begin{align*}\mathbf{G}_{\text {\textbf{Inf}}}=
    \begin{Vmatrix}
        c^{(\tau-1)}_{0} & c^{(\tau-1)}_{1} & \ldots\ldots & c^{(\tau-1)}_{\tau-2} & c^{(\tau-1)}_{\tau-1} \\
        c^{(\tau-2)}_{0} & c^{(\tau-2)}_{1} & \ldots\ldots & c^{(\tau-2)}_{\tau-2} & c^{(\tau-2)}_{\tau-1} \\
       \hdotsfor[2]{5} \\
        c^{(1)}_{0} & c^{(1)}_{1} & \ldots\ldots & c^{(1)}_{\tau-2} & c^{(1)}_{\tau-1} \\
        c^{(0)}_{0} & c^{(0)}_{1} & \ldots\ldots & c^{(0)}_{\tau-2} & c^{(0)}_{\tau-1}
    \end{Vmatrix}^{\top}.
\end{align*}

Thus we obtain the $q$-th block of the PRS:
\begin{align*}
    \textbf{X}_{q}=\mathbf{G}_{\text {\textbf{Inf}}}\cdot \mathbf{X}_{q-1},
\end{align*}
where
\begin{align*}\mathbf{X}_{q}&=
    \begin{bmatrix}
        x_{q,\ \tau-1} & x_{q,\ \tau-2} & \ldots & x_{q,\ 1} & x_{q,\ 0}
    \end{bmatrix}^{\top},\\
\mathbf{X}_{q-1}&=
    \begin{bmatrix}
        x_{q-1,\ \tau-1} & x_{q-1,\ \tau-2} & \ldots & x_{q-1,\ 1} & x_{q-1,\ 0}
    \end{bmatrix}^{\top}.
\end{align*}

To create the conditions for the application of separable linear redundant
code will get form a matrix $\mathbf{G}_{\text
{\textbf{Gen}}}$, consisting of the information and the check matrix by adding
(3) validation expressions:
\begin{align*}
    \begin{cases}
        x_{q,\ \tau-1}=c^{(\tau-1)}_{0}x_{q-1,\ 0}\oplus c^{(\tau-1)}_{1}x_{q-1,\ 1}\oplus \ldots \oplus c^{(\tau-1)}_{\tau-2}x_{q-1,\ \tau-2}\oplus c^{(\tau-1)}_{\tau-1}x_{q-1,\ \tau-1},\\
        x_{q,\ \tau-2}=c^{(\tau-2)}_{0}x_{q-1,\ 0}\oplus c^{(\tau-2)}_{1}x_{q-1, 1}\oplus \ldots \oplus c^{(\tau-2)}_{\tau-2}x_{q-1,\ \tau-2}\oplus c^{(\tau-2)}_{\tau-1}x_{q-1,\ \tau-1},\\
        \cdots\cdots\cdots\cdots\cdots\cdots\cdots\cdots\cdots\cdots\cdots\cdots\cdots\cdots\cdots\cdots\cdots\cdots\cdots\cdots\cdots\cdots
        \cdots\cdots\cdots\cdots\\
        x_{q,\ 1}=c^{(1)}_{0}x_{q-1,\ 0}\oplus c^{(1)}_{1}x_{q-1,\ 1}\oplus \ldots \oplus c^{(1)}_{\tau-2}x_{q-1,\ \tau-2}\oplus c^{(1)}_{\tau-1}x_{q-1,\ \tau-1},\\
        x_{q,\ 0}=c^{(0)}_{0}x_{q-1,\ 0}\oplus c^{(0)}_{1}x_{q-1,\ 1}\oplus \ldots \oplus c^{(0)}_{\tau-2}x_{q-1,\ \tau-2}\oplus c^{(0)}_{\tau-1}x_{q-1,\ \tau-1},\\
        x^{*}_{q,\ r-1}=a^{(r-1)}_{0}x_{q-1,\ 0}\oplus a^{(r-1)}_{1}x_{q-1,\ 1}\oplus \ldots \oplus a^{(r-1)}_{\tau-2}x_{q-1,\ \tau-2}\oplus a^{(r-1)}_{\tau-1}x_{q-1,\ r-1},\\
        \cdots\cdots\cdots\cdots\cdots\cdots\cdots\cdots\cdots\cdots\cdots\cdots\cdots\cdots\cdots\cdots\cdots\cdots\cdots\cdots\cdots\cdots
        \cdots\cdots\cdots\cdots\\
        x^{*}_{q,\ 0}=a^{(0)}_{0}x_{q-1,\ 0}\oplus a^{(0)}_{1}x_{q-1,\ 1}\oplus \ldots \oplus a^{(0)}_{\tau-2}x_{q-1,\ \tau-2}\oplus a^{(0)}_{\tau-1}x_{q-1,\ \tau-1},
    \end{cases}
\end{align*}
where $r$~--- the number of redundant symbols used linear code, $a^{(j)}_{i}\in\{0,\ 1\}$  $(i=0,\ 1,\ \ldots,\ \tau-1;\ \   j=0,\ \ldots,\ r-1)$.

A generator matrix takes the form:

\begin{align*}\mathbf{G}_{\text {\textbf{Gen}}}=
    \begin{Vmatrix}
        c^{(\tau-1)}_{0} & c^{(\tau-1)}_{1} & \ldots\ldots & c^{(\tau-1)}_{\tau-2} & c^{(\tau-1)}_{\tau-1} \\
        c^{(\tau-2)}_{0} & c^{(\tau-2)}_{1} & \ldots\ldots & c^{(\tau-2)}_{\tau-2} & c^{(\tau-2)}_{\tau-1} \\
        \hdotsfor[2]{5} \\
        c^{(1)}_{0} & c^{(1)}_{1} & \ldots\ldots & c^{(1)}_{\tau-2} & c^{(1)}_{\tau-1} \\
        c^{(0)}_{0} & c^{(0)}_{1} & \ldots\ldots & c^{(0)}_{\tau-2} & c^{(0)}_{\tau-1} \\
        a^{(r-1)}_{0} & a^{(r-1)}_{1} & \ldots\ldots & a^{(r-1)}_{\tau-2} & a^{(r-1)}_{\tau-1} \\
        \hdotsfor[2]{5} \\
        a^{(0)}_{0} & a^{(0)}_{1} & \ldots\ldots & a^{(0)}_{\tau-2} & a^{(0)}_{\tau-1}
    \end{Vmatrix}^{\top}.
\end{align*}

Then the $q$-th block of the PRS with the control numbers (linear block code):
\begin{align*}\textbf{X}^{*}_{q}=
    \begin{bmatrix}
        x_{q,\ \tau-1} & x_{q,\ \tau-2} & \ldots & x_{q,\ 1} & x_{q,\ 0} & x^{*}_{q,\ r-1} & \ldots & x^{*}_{q,\ 0}
    \end{bmatrix}^{\top}
\end{align*}
is calculated by:
\begin{align*}
    \mathbf{X}^{*}_{q}=\mathbf{G}_{\text {\textbf{Gen}}}\cdot \mathbf{X}_{q-1}.
\end{align*}

Procedure error-correcting decoding is performed using the known rules [4].
The application of linear redundant codes and methods ``hot'' standby is not
the only option for the implementation of functional diagnostics and fault
tolerance of digital devices. Example graph parallel generation elements PRS error control computations is shown in Fig.~3.

\begin{figure}[h]
    \begin{center}
    \resizebox{1.0\linewidth}{!}{
\colorbox[gray]{.92}{
$$
\xymatrix{
\text{\textsf{Output $(q)$  PRS block} }  &    x_{q,\ 3}                                           & x_{q,\ 2}                                            & x_{q,\ 1}                                                                & x_{q,\ 0}                                                    &  x_{q,\ 0}^* \\
\text{\textsf{Step}\ $q$}             & \bigoplus   \ar@{->}[u]                          & \bigoplus  \ar@{->}[u]                           & \bigoplus   \ar@{->}[u]                                              & \bigoplus  \ar@{->}[u]                               &    \ar@{-->}[u]  \\
\text{ \textsf{Input $(q-1)$  PRS block} } &   x_{q-1,\ 3} \ar@{->}[u]\ar@{->}[ur]    &  x_{q-1,\ 2}\ar@{->}[u]\ar@{->}[ur]   & x_{q-1,\ 1}\ar@{->}[u]\ar@{->}[ur]\ar@{->}[ull]\ar@{--}[urr]      &  x_{q-1,\ 0} \ar@{->}[u]  \ar@{->}[ulll]  & x_{q-1,\ 0}^*
 }
$$
}
}
  \end{center}
   \begin{center}
   a)
   \end{center}
  \begin{center}
    \resizebox{0.7\linewidth}{!}{
\colorbox[gray]{.92}{
$$
\xymatrix{
                                             &  1                                                         & 1                                                          & 0                                                                               & 0                                                       & 0  \\
\text{\textsf{Step\ 3}}             & \bigoplus   \ar@{->}[u]                          & \bigoplus  \ar@{->}[u]                           & \bigoplus   \ar@{->}[u]                                              & \bigoplus  \ar@{->}[u]   &   \ar@{-->}[u]                     \\
                                             &1 \ar@{->}[u]\ar@{->}[ur]   &  0   \ar@{->}[u]\ar@{->}[ur]   & 0   \ar@{--}[urr]   \ar@{->}[u]\ar@{->}[ur]\ar@{->}[ull]     &  0  \ar@{->}[u]  \ar@{->}[ulll]           & 1   \\
\text{\textsf{Step\ 2}}             & \bigoplus   \ar@{->}[u]                          & \bigoplus  \ar@{->}[u]                           & \bigoplus   \ar@{->}[u]                                              & \bigoplus  \ar@{->}[u]   &    \ar@{-->}[u]  \\
                                             & 1  \ar@{->}[u]\ar@{->}[ur]   & 1  \ar@{->}[u]\ar@{->}[ur]   & 1  \ar@{--}[urr]  \ar@{->}[u]\ar@{->}[ur]\ar@{->}[ull]     &  1   \ar@{->}[u]  \ar@{->}[ulll]           & 0   \\
\text{\textsf{Step\ 1}}             & \bigoplus   \ar@{->}[u]                          & \bigoplus  \ar@{->}[u]                           & \bigoplus   \ar@{->}[u]                                              & \bigoplus  \ar@{->}[u]   &  \ar@{-->}[u]    \\
\text{\textsf{Entry conditions}} & 0   \ar@{->}[u]\ar@{->}[ur]  &  1  \ar@{->}[u]\ar@{->}[ur]   & 0  \ar@{--}[urr] \ar@{->}[u]\ar@{->}[ur]\ar@{->}[ull]     & 1    \ar@{->}[u]  \ar@{->}[ulll] &    0
 }
$$
}
}
  \end{center}
  \begin{center}
   b)
   \end{center}
    \caption{a) Example graph parallel generation elements PRS (the characteristic equation: $x_{p+4}=x_{p+1}\oplus x_{p}$) error control computations (parity control); b)  Numerical example}
\end{figure}

Important advantages for these purposes have
redundant arithmetic codes, in particular, so-called $AN$-codes and
residue number systems  (RNS) codes. The application of these codes to monitor
logical data types and fault tolerance implementing devices became possible
with the introduction of logical operations arithmetic expressions [11], in
particular linear numerical polynomials (LNP) and modular forms [12].

\section{Error control operation of the PRS generators,
based on ``arithmetization'' logical account}

At the end of the last century there was formed a new direction parallel
logic computation by the arithmetic (numeric) polynomials [11]. In particular
received position ``Modular arithmetic parallel logic computation'' of the unification
of the theoretical foundations of RNS [13, 14, 15] and theoretical foundations of
parallel logic computation by the arithmetic of polynomials. The objective of the
Association is to use advantages of RNS, i.e. parallelization arithmetic, error
control calculations [16] in real time and ensure high availability of computing
equipment, in the field of parallel logical account. In the following, these
provisions were developed in various aspects, in particular, towards the
implementation of cryptographic functions [17, 18]. In particular, it was
considered parallel generators PRS based, in General, nonlinear (canonical)
arithmetic polynomials. Using use of LNP proposed by prof. V.D. Malyugin [11]
for the construction of parallel generators PRS possible to reduce the maximum
length of realizing polynomial to a value of $n+1$, where $n$~--- number of
arguments of a Boolean function implemented [18]. In this paper, this method is
used as the basis for the construction of safe (self-checking, fault-tolerant)
generators on the basis of the excess bandwidth RNS.

It is known [19] that the $q$-th block of land PRS can be represented by a single LNP.
The system of characteristic equations (3) must submit, as a system of Boolean functions,
which in turn must be converted into a system:
\begin{align*}
    \begin{cases}
        L_{\tau-1}(\textbf{X}_{q-1})=g^{(\tau-1)}_{1}x_{q-1,\ 0}+g^{(\tau-1)}_{2}x_{q-1,\ 1}+\ldots+g^{(\tau-1)}_{\tau}x_{q-1,\ \tau-1},\\
        L_{\tau-2}(\textbf{X}_{q-1})=g^{(\tau-2)}_{1}x_{q-1,\ 0}+g^{(\tau-2)}_{2}x_{q-1,\ 1}+\ldots+g^{(\tau-2)}_{\tau}x_{q-1,\ \tau-1},\\
        \cdots\cdots\cdots\cdots\cdots\cdots\cdots\cdots\cdots\cdots\cdots\cdots\cdots\cdots\cdots\cdots\cdots\cdots\cdots\cdots\cdots\cdots\\
        L_{0}(\textbf{X}_{q-1})=g^{(0)}_{1}x_{q-1,\ 0}+g^{(0)}_{2}x_{q-1,\ 1}+\ldots+g^{(0)}_{\tau}x_{q-1,\ \tau-1},
    \end{cases}
\end{align*}
where $g_{j}^{(i)}$  (here and then) takes the value ``0'' or ``1'' depending on the entry in the
$i$-th LNP $x_{q-1,\ j}$; $i,\ j= 0,\ 1,\ \ldots,\ \tau-1$.

The result of the calculation of $i$-LNP system appears to be a binary word of length
$l_{i}=\lfloor \log(\sum\limits^{0}_{j=\tau-1}g^{(i)}_{j})\rfloor+1$,
where $\lfloor a\rfloor$~--- the largest integer. Calculated total LNP:
\begin{align*}
    \begin{split}
        L(\textbf{X}_{q-1})&=L_{\tau-1}(\textbf{X}_{q-1})+2^{\gamma_{1}}L_{\tau-2}(\textbf{X}_{q-1})+
        \ldots +2^{\gamma_{\tau-1}}L_{0}(\textbf{X}_{q-1})= \\
        &= g^{(\tau-1)}_{1}x_{q-1,\ 0}+g^{(\tau-1)}_{2}x_{q-1,\ 1}+\ldots+g^{(\tau-1)}_{\tau}x_{q-1,\ \tau-1}+ \\
        &+2^{\gamma_{1}}(g^{(\tau-2)}_{1}x_{q-1,\ 0}+g^{(\tau-2)}_{2}x_{q-1,\ 1}+\ldots+g^{(\tau-2)}_{\tau}x_{q-1,\ \tau-1})+\ldots \\
        &\ldots+2^{\gamma_{\tau-1}}(g^{(0)}_{1}x_{q-1,\ 0}+g^{(0)}_{2}x_{q-1,\ 1}+\ldots+g^{(0)}_{\tau}x_{q-1,\ \tau-1})= \\
        &=h_{1}x_{q-1,\ 0}+h_{2}x_{q-1,\ 1}+\ldots+h_{\tau}x_{q-1,\ \tau-1},
    \end{split}
\end{align*}
where $\gamma_{k}=\sum\limits_{i=0}^{k-1}(l_{i}+1)$, $k=1,\ 2,\ \ldots,\ \tau-1$; $h_{j}\in Z$, or
\begin{align}
    L(\textbf{X}_{q-1})=\sum_{i=1}^{\tau}h_{i}x_{q-1,\ i-1}.
\end{align}

The final result is formed by implementing operator masking $\Xi^{\varphi}\{U\}$,
which is used to determine the values of the $\varphi$-th Boolean function representation
$U=(b_{v}\ldots b_{\varphi}\ldots b_{2}b_{1})_{2}$ (record
$(\ldots)_{2}$ means representing a non-negative $U$
in a binary number), that is $\Xi^{\varphi}\{U\}=b_{\varphi}$.
Count calculation LNP (4) is represented in figure~3.

In RNS a nonnegative coefficient LNP (4) $h_{j}$ is uniquely represented by a
set of residues on the grounds RNS ($m_{1},\ m_{2},\
\ldots,\ m_{n}<m_{n+1}<\ldots<m_{k}$~--- pairwise simple):
\begin{align}
    h_{j}=(\alpha_{1},\ \alpha_{2},\ \ldots,\ \alpha_{n},\ \alpha_{n+1},\ \ldots,\ \alpha_{k})_{\text{MA}},
\end{align}
where $\alpha_{t}=|h_{j}|_{m_{t}}$; $t=1,\ 2,\ \ldots,\ n,\ \ldots,\ k$;
$|\centerdot|_{m}$~--- the smallest non-negative deduction number $\centerdot$ on the modulo $m$.
Operating range $M_{n}=m_{1}m_{2}\ldots m_{n}$ must meet
$M_{n}>2^{s}$, where $s=\sum\limits_{1\leq\varepsilon\leq\tau}l_{\varepsilon}$~---
the number of binary bits required to represent the result of a calculation LNP (4).

The remains  $\alpha_{1},\ \alpha_{2},\ \ldots,\ \alpha_{n}$ are informational, and
$\alpha_{n+1},\ \ldots,\ \alpha_{k}$~---
are control. RNS in this case is called the extended and covers the complete set of
states represented all $k$ residues. This area is full range RNS $[0,\
M_{k})$, where $M_{k}=m_{1}m_{2}\ldots m_{n}m_{n+1}\ldots m_{k}$, and consists of the operating range
$[0,\ M_{n})$, defined information bases RNS, and range identified redundant bases $[M_{n},\ M_{k})$,
unacceptable region for the results of a calculation. This means that operations on numbers $h_{j}$
are in the range  $[0,\ M_{k})$. Therefore, if the result of the operation RNS beyond $M_{n}$,
it should output error calculation.

Consider RNS specified grounds $m_{1},\ m_{2},\ \ldots,\ m_{n},\
m_{n+1}$. Each coefficient LNP $h_{j}$ can be written as (5) and get redundant code RNS
represented by the LNP system:
\begin{align}
    \begin{cases}
        U^{(1)}=L^{(1)}(\textbf{X}_{q-1})=\alpha^{(1)}_{1}x_{q-1, 0}+\alpha^{(1)}_{2}x_{q-1, 1}+\ldots+\alpha^{(1)}_{\tau}x_{q-1, \tau-1},\\
        U^{(2)}=L^{(2)}(\textbf{X}_{q-1})=\alpha^{(2)}_{1}x_{q-1, 0}+\alpha^{(2)}_{2}x_{q-1, 1}+\ldots+\alpha^{(2)}_{\tau}x_{q-1, \tau-1},\\
        \cdots\cdots\cdots\cdots\cdots\cdots\cdots\cdots\cdots\cdots\cdots\cdots\cdots\cdots\cdots\cdots\cdots\cdots\cdots\cdots\cdots\cdots\\
        U^{(n)}=L^{(n)}(\textbf{X}_{q-1})=\alpha^{(n)}_{1}x_{q-1, 0}+\alpha^{(n)}_{2}x_{q-1, 1}+\ldots+\alpha^{(n)}_{\tau}x_{q-1, \tau-1},\\
        U^{(n+1)}=L^{(n+1)}(\textbf{X}_{q-1})=\alpha^{(n+1)}_{1}x_{q-1, 0}+\alpha^{(n+1)}_{2}x_{q-1, 1}+\ldots\\
        \ldots+ \alpha^{(n+1)}_{\tau}x_{q-1, \tau-1}.
    \end{cases}
\end{align}

Substituting in (6) values of RNS residue on the appropriate grounds for each coefficient
(4) and the values of the variables
$x_{q-1,\ 0},\ \ldots,\ x_{q-1,\ \tau-1}$, get the values of LNP system (6), where $U^{(1)},\ U^{(2)},\ \ldots,\ U^{(n)},\
U^{(n+1)}$~--- nonnegative integer. In accordance with the Chinese remainder theorem solve the system of equations:
\begin{align}
 \begin{cases}
    U^{*}=|U^{(1)}|_{m_{1}},\\
     U^{*}=|U^{(2)}|_{m_{2}},\\
      \ldots\ldots\ldots\ldots\ldots\ldots\\
       U^{*}=|U^{(n)}|_{m_{n}},\\
        U^{*}=|U^{(n+1)}|_{m_{n+1}}.
    \end{cases}
\end{align}

Since $m_{1},\ m_{2},\ \ldots,\ m_{n},\ m_{n+1}$ are pairwise prime, then
the only solution of (7) gives the expression:
\begin{align}
    U^{*}=\Biggl|\sum_{s=1}^{n+1}M_{s,\ n+1}\mu_{s,\ n+1}U^{(s)}\Biggl|_{M_{n+1}},
\end{align}
where $M_{s,\ n+1}=\dfrac{M_{n+1}}{m_{s}}$,\  $\mu_{s,\ n+1}=|M^{-1}_{s,\ n+1}|_{m_{s}}$,\
$M_{n+1}=\prod\limits_{s=1}^{n+1}m_{s}$.

Graph parallel generation PRS based on (8) is shown in Fig. 4. The occurrence of the result of the calculation (8) in the range (control expression):
\begin{align*}
    0\leq U^{*}<M_{n},
\end{align*}
means the absence of detectable errors of calculations.

\begin{figure}[h]
    \begin{center}
    \resizebox{0.99\linewidth}{!}{
\colorbox[gray]{.92}{
$$
\xymatrix{
&x_{q,\ \tau -1}    &   x_{q,\ \tau -2}  &  \ldots\ldots  &   x_{q-1,\ 0}    \\
   &\Xi^{(\tau -1)} \ar@{->}[u] &\Xi^{(\tau -2)} \ar@{->}[u] &\ldots\ldots&\Xi^{(0)}\ar@{->}[u]  \\
   &\ar@{->}[u]& \ar@{->}[u] \ar@{-}[r]\ar@{-}[l]&U^*\ar@{-}[r]&\ar@{->}[u]\\
&&&\text{\textsf{\textbf{CRT}}}\ar@{-}[u]&&\\
&&&&&\\
L^{(1)}(\mathbf{X}_{q-1}) \ar@{->}[uurrr]&L^{(2)}(\mathbf{X}_{q-1}) \ar@{->}[uurr]   &\cdots\cdots   &    L^{(n)}(\mathbf{X}_{q-1}) \ar@{->}[uu]   &\cdots\cdots& L^{(n+k)}(\mathbf{X}_{q-1})\ar@{-->}[uull]\\
&&&&&\\
&x_{q-1,\ \tau -1} \ar@{->}[uu] \ar@{->}[uul]\ar@{->}[uurr]\ar@{->}[uurrrr]&x_{q-1,\ \tau -2} \ar@{->}[uull]\ar@{->}[uul]\ar@{->}[uur]\ar@{->}[uurrr]&\ldots\ldots& x_{q-1,\ 0} \ar@{->}[uur]\ar@{->}[uul]\ar@{->}[uulll]\ar@{->}[uullll]
 }
$$
}
}
  \end{center}
   \caption{Graph of parallel generation PRS based on the Chinese remainder theorem (CRT)}
        \end{figure}
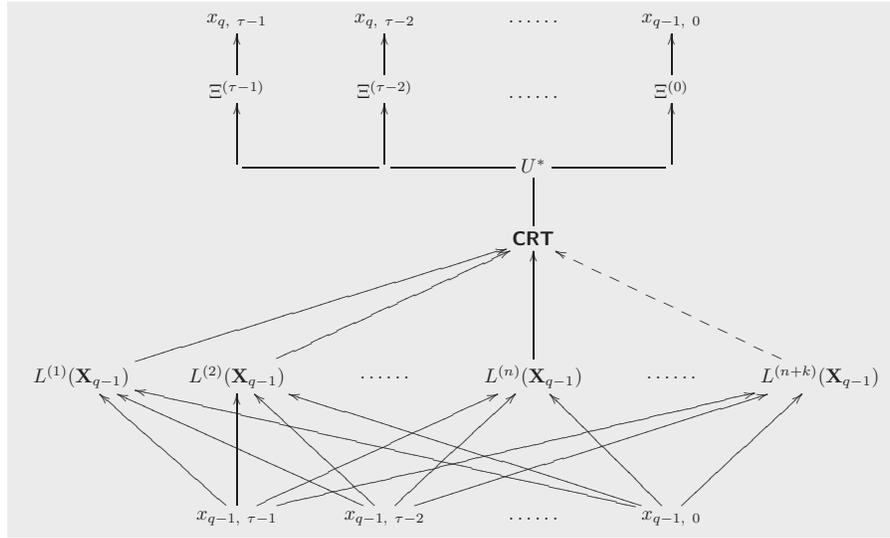

\section{Reconfiguration of equipment}

Restore reliable operation of the generator of the PRS in the case of long-term
failure is possible by correcting an error or reconfiguration of equipment
generator (active redundancy). The first option is unacceptable because it
does not guarantee no penetration of undetectable errors in the result of
the encryption. By methods of modular redundant coding is made possible to
apply a variant of the reconfiguration of the equipment by excluding from
the operation of the failed equipment.

After localization of the faulty equipment~--- for example~--- a single channel
operation RNS, the reconfiguration operation is performed by the calculation $U^{*}$
from the system:
\begin{align*}
    \begin{cases}
    U^{*}=|U^{(1)}|_{m_{1}},\\
     \ldots\ldots\ldots\ldots\ldots\ldots\\
      U^{*}=|U^{(n)}|_{m_{n}},\\
       U^{*}=|U^{(n+1)}|_{m_{n+1}},\\
        U^{*}=|U^{(n+2)}|_{m_{n+2}}
    \end{cases}
\end{align*}
on the ``right'' reasons of residue number systems:
\begin{align*}
    U^{*}=|\widetilde{U}^{(1)}B_{1,\ j}+\widetilde{U}^{(2)}B_{2,\ j}+\ldots+\widetilde{U}^{(n+2)}B_{n+2,\ j}|_{M_{j}},
\end{align*}
where $\widetilde{U}^{(i)}$~--- wrong balance; $B_{i,\ j}$~---
orthogonal bases; $i,\ j=1,\ 2,  \ldots,\  n+2$; $i\neq j$; $B_{i,\
j}=\dfrac{M_{j}\mu_{i,\  j}}{m_{i}}$;
$M_{j}=\dfrac{M_{n+2}}{m_{j}}$; $\mu_{i,\  j}$ is calculated from the comparison:
$\dfrac{M_{j}\mu_{i,\  j}}{m_{i}}\equiv1 \pmod {m_{i}}$.
Compiled table~1 contains the values of the orthogonal bases and modules
of the system for the occurrence of a single error for each base RNS.
\begin{table}[h]
\caption{Calculation table orthogonally bases and modules RNS}
    \begin{tabular}{c|cccc|c}
    \hline\noalign{\smallskip}
        $j$ & $B_{1,\ j}$ & $B_{2,\ j}$ &$\cdots$ & $B_{n+2,\ j}$ & $M_{j}$ \\
        \noalign{\smallskip}\hline\noalign{\smallskip}
               $1$ & 0 & $\dfrac{M_{1}\mu_{2,\ 1}}{m_{2}}$ & $\cdots$ &$\dfrac{M_{1}\mu_{n+2,\ 1}}{m_{n+2}}$ & $m_{2}m_{3}\ldots m_{n+2}$ \\
                $2$ & $\dfrac{M_{2}\mu_{1,\ 2}}{m_{1}}$ & 0 & $\cdots$ & $\dfrac{M_{2}\mu_{n+2,\ 2}}{m_{n+2}}$ & $m_{1}m_{3}\ldots m_{n+2}$ \\
               $\cdots\cdots$ & $\cdots\cdots\cdots$ & $\cdots\cdots\cdots$ &  $\cdots$   & $\cdots\cdots\cdots$ & $\cdots\cdots\cdots\cdots\cdots$ \\
               $n+2$ & $\dfrac{M_{n+2}\mu_{1,\ n+2}}{m_{1}}$ &  $\dfrac{M_{n+2}\mu_{2,\ n+2}}{m_{2}}$ & $\cdots$ & 0 & $m_{1}m_{2}\ldots m_{n+1}$ \\
               \noalign{\smallskip}\hline
           \end{tabular}
\end{table}

\section{Conclusion}
It is known that the use of RNS already with two redundant bases allows
us to provide a level of fault tolerance modular transmitter exceeds the
tolerance provided by the method of rorovana equipment. This redundant
hardware costs are reduced from 200\% (triple) up to 30-40\% (when using RNS)
[20]. At the same time it should be noted that the amount of hardware, PRS
generator operating in accordance obtained by the method, may exceed the
hardware failover LFSR, built in accordance with traditional solutions.
So you should made a fundamentally new level of functional flexibility
of the designed generator PRS is able to implement and many other cryptographic
functions, time-varying, without rebuilding the structure. This allows for the
implementation of the device not only programmable logic integrated circuit,
but also high-tech large custom integrated circuits, in particular used for
the implementation of number theoretic transformations in the field of digital
signal processing.

The implementation of the PRS generators using LNP and redundant RNS allows
to obtain a new class of solutions aimed at the safe implementation of the
logical cryptographic functions, in particular parallel generators PRS. This
is provided as a functional control equipment (in real time), and its fault
tolerance through reconfiguration of the structure of the evaluator in the
process of its degradation. Classic LFSR considered in the present work, is
the basis and more complex, for example, combining generators PRS. Use for
the implementation of the PRS generator modular arithmetic provides the
possibility of applying the proposed solutions in the hybrid cryptosystems
(including asymmetric) [18]. When this arithmetic calculator that supports
the implementation of asymmetric cryptographic algorithms may be used to
implement systems of Boolean functions (elements PRS).

\end{document}